# *MotiBo*: The Impact of Interactive Digital Storytelling Robots on Student Motivation through Self-Determination Theory

Ka Yan FUNG
Institute of Special Needs and Inclusive Education
The Education University of Hong Kong
Hong Kong SAR, China
fkayan@eduhk.hk

Tze Leung Rick LUI
Institute of Special Needs and Inclusive Education
The Education University of Hong Kong
Hong Kong SAR, China
rtllui@eduhk.hk

Yuxing TAO
Institute of Special Needs and Inclusive Education
The Education University of Hong Kong
Hong Kong SAR, China
ytao@eduhk.hk

Kuen Fung SIN
Institute of Special Needs and Inclusive Education
The Education University of Hong Kong
Hong Kong SAR, China
kfsin@eduhk.hk

## ABSTRACT

Creativity is increasingly recognized as an important skill in education, and storytelling can enhance motivation and engagement among students. However, conventional storytelling methods often lack the interactive elements necessary to engage students. To this end, this study examines the impact of an interactive digital storytelling system incorporating a human-like robot on student engagement and creativity. The study aims to compare engagement levels across three modalities: paper-based, PowerPoint, and robot-assisted storytelling, *MotiBo*. Utilizing a quasi-experimental design, this work involves three groups of students who interact with the storytelling system over a five-day learning. Findings reveal that students using *MotiBo* exhibit statistically significant improvement in behavioural and cognitive engagement compared to those using traditional methods. These results suggest that the integration of novel technologies can effectively enhance the learning experience, ultimately promoting creativity and self-learning ability in educational settings. Future research will investigate the long-term effects of these technologies on learning outcomes and explore their potential for broader applications in diverse educational contexts.

## 1 Introduction

***Creativity*** is increasingly recognized as a crucial skill in education, as it can enhance students' critical thinking and problem-solving ability [8]. Storytelling is an effective catalyst for nurturing creativity and cognitive development in students [1, 5]. With the advancement of novel technologies, **digital storytelling** has transcended its traditional boundaries, offering new modalities for engagement and interaction that could potentially amplify educational benefits [13]. To further enhance students' learning motivation, **interactive social robots** can play a pivotal role [3, 4]. By incorporating multisensory supports, such as animations, **instant responses**, and movements, social robots can effectively engage students and capture their attention [6]. Despite the recognized potential benefits of digital storytelling to enhance creativity, there remains a significant gap in understanding how different modalities of storytelling delivery impact student engagement. While traditional methods, such as paper and PowerPoint (PPT)-based learning, have been widely used in educational contexts [9, 14], they often lack the interactive and immersive qualities that can captivate students' interest. This study aims to address this gap by comparing engagement levels among students using **paper-based, PPT-based, and robot-assisted** learning experiences. We can gain valuable insights into the effectiveness of these different approaches. Understanding these differences is essential for educators seeking to optimize learning environments and foster creativity in their students.

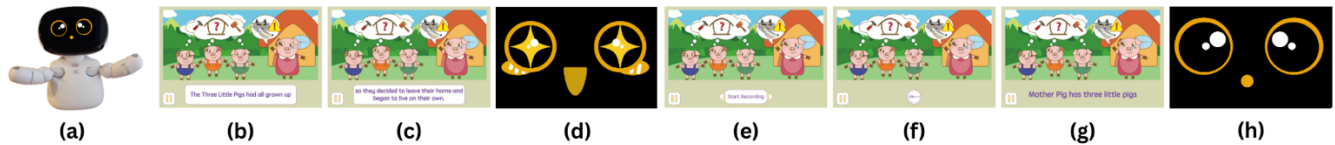

Figure 1: The overview of the robot, UI and flow of digital storytelling: (a) is the robot, *Kebby*; (b) & (c) read students the story with sub-title and a multiple listening function; (d) asks students questions; (e) allows students to press for an answer; (f) notifies students ASR processing; (g) repeats the answer; (h) praises students' learning efforts.

## 2  System Design

The study employs a two-stage approach, including system development and empirical research. In the following section, we will elaborate on the system development process and the pilot study.

### 2.1  Game Prototype

In this work, we designed an interactive digital storytelling system, *MotiBo* (Figure 1), that features five well-known stories (e.g., "Three Little Pigs") presented on a human-like robot, *Kebby* [10].

**Game Flow.** *MotiBo* begins when students log in, and the storytelling procedure is outlined as follows. First, to ensure fairness, students are required to experience the stories in sequence. Second, students listen to each story and answer related questions, with a maximum of three attempts allowed for each question. Each attempt is limited to 10 seconds. Third, *MotiBo* provides instant feedback to students using Automatic Speech Recognition (ASR) technology, indicating the correctness of their answers while also delivering motivational phrases and animations to encourage engagement.

**Stories.** *MotiBo* consists of five famous stories, such as "Snow White". Each story comprises five pictures. Each picture includes two to four sentences that convey the narrative content.

**Robot.** *Kebby* is equipped with a movable head, two hands, and three wheels. It is operated by seven motors that control various body parts, including the neck, fists, elbows, and shoulders. *Kebby* can be programmed to perform a wide range of movements and gestures that mimic human-like actions.

**User Interface Design.** *MotiBo* integrates multisensory functions and interactive design features to ensure active student participation. It provides feedback on students' responses, whether correct or incorrect, and can deliver instant responses.

### 2.2  Efficacy Experiment

Three groups of students participated in the study to compare their engagement levels across three modalities (same story materials): paper, PPT, and robot-assisted storytelling. The *MotiBo* system supports three languages: Traditional Chinese (Cantonese), Simplified Chinese (Mandarin), and English, allowing students to choose their preferred language.

**Participants. Paper-based** group consists of eight students (4 female and 4 male) with average age 7.73 year-old (SD= 0.58 year-old). **PPT** group consists of eight students (4 female and 4 male) with average age 7.96 year-old (SD= 0.84 year-old). ***MotiBo*** group consists of eight students (4 female and 4 male) with average age 8.02 year-old (SD= 0.60 year-old).

**Procedure.** On Day 1, students began with a pre-questionnaire before starting the storytelling activities. A total of five stories were presented, with one story each day, and each story took approximately 20 minutes to complete. An instructor supported each student by either reading the story to them or providing technical assistance. On Day 5, students conducted a post-questionnaire and an interview.

**Self-determination Theory and Interview.** Self-Determination Theory (SDT) is a psychological framework that promotes intrinsic motivation and engagement in learning [11]. A total of 14 questions were developed, covering behaviour, emotion, cognition, and intrinsic motivation, with three questions dedicated to each area. An example question is, "I find learning a story with [paper/PPT/robot] fun." Additionally, semi-structured interviews were conduct gather student feedback, with a sample question being, "How to improve your engagement when learning a story with [paper/PPT/robot]?"

**Inclusion Criteria.** Eligibility for students' participation required enrolling in Grades 1 to 3 and having no medical or physical disabilities that could hinder interaction with the robot or affect communication abilities. Additionally, all participants had prior experience using digital tools, such as tablets. Informed consent was obtained from the parents of all students before the experiment commenced. Participation was entirely voluntary and contingent upon parental consent. The experimental protocol received approval from the University Institutional Review Board (IRB). No remuneration was offered to participants.

## 3 Efficacy Verification

This section reveals data analysis on engagement and motivation in three modalities.

**Behavioural Engagement.** The robot group exhibited a statistically significant improvement, $p<.05$ (pre-test: $\bar{M}$ = 70.83, $\sigma$ = 18.15; post-test: $\bar{M}$= 90.00, $\sigma$ = 9.43; +27.06%). The PPT group demonstrated improvement, $p$ =.09 (pre-test: $\bar{M}$= 79.17, $\sigma$ =4.71; post-test: $\bar{M}$= 91.67, $\sigma$ =11.51; +15.79%). The paper group also shown enhancement, $p$ =.19 (pre-test: $\bar{M}$= 72.50, $\sigma$ = 9.72; post-test: $\bar{M}$= 80.83, $\sigma$ = 14.00; +11.49%).

**Emotional Engagement.** The robot group demonstrated improvement, $p$ =.06 (pre-test: $\bar{M}$= 75.83, $\sigma$ =22.80; post-test: $\bar{M}$= 93.33, $\sigma$ =7.97; +23.08%). The PPT group shown an increase, $p$ =.11 (pre-test: $\bar{M}$=85.00, $\sigma$ =19.41; post-test: $\bar{M}$= 92.50, $\sigma$ =17.41; +8.82%). The paper also exhibited increased, $p$ =.38 (pre-test: $\bar{M}$ = 75.00, $\sigma$ =13.21; post-test: $\bar{M}$ = 81.67, $\sigma$ =15.84; +8.89%).

**Cognitive Engagement.** The robot group displayed a statistically significant improvement, $p<.05$ (pre-test: $\bar{M}$ = 66.67, $\sigma$ =18.86; post-test: $\bar{M}$ = 86.67, $\sigma$ =10.08; +30.00%). The PPT group reflected advancement, $p$ =.23 (pre-test: $\bar{M}$ = 81.67, $\sigma$ =10.54; post-test: $\bar{M}$= 89.17, $\sigma$ =13.30; +9.18%). The paper group indicated positive improvement, $p$ =.28 (pre-test: $\bar{M}$ = 69.17, $\sigma$ =14.23; post-test: $\bar{M}$ = 76.67, $\sigma$ =12.34; +10.84%).

**Intrinsic Motivation.** The robot group manifested improvement, $p$ =.07 (pre-test: $\bar{M}$= 78.33, $\sigma$ =21.31; post-test: $\bar{M}$ = 94.17, $\sigma$ =7.51; +20.21%). The PPT group noted a boost, $p$ =.06 (pre-test: $\bar{M}$ = 80.83, $\sigma$ =13.54; post-test: $\bar{M}$ = 94.17, $\sigma$ =12.05; +16.49%). The paper group slightly increased, $p$ =.67 (pre-test: $\bar{M}$ = 85.00, $\sigma$ =12.22; post-test: $\bar{M}$ = 88.33, $\sigma$ =18.08; +3.92%).

## 4 Discussion

Robot-assisted learning can increase students' learning interest. When the instructor asked if they preferred robot-assisted storytelling over traditional methods, Student A replied, "Of course, I prefer this little robot. It can dance. How can the teacher dance like that for me?" This approach also enhances students' enthusiasm, and a strong learning initiative is essential for improving cognitive engagement [2]. For example, after the storytelling practice, Student B asked, "Can I listen to this story again?" Students in the *MotiBo* group were more excited and actively sought additional storytelling practice compared to those in other groups.

Language proficiency is positively correlated with students' creativity [7]. *MotiBo* encourages students to practice storytelling in their preferred language, allowing them to express their creativity fully. Smakman et al. [12] found that students often trust robots more, even upon first meeting them. For example, Student C was reluctant to speak to the instructor but readily answered the robot's questions, suggesting that robots can act as more equal communication partners.

## 5 Conclusion and Future Work

To conclude, storytelling has a positive impact in schools. The three modalities show enhancement based on SDT. Notably, *MotiBo* effectively improves students' behavioural engagement and intrinsic motivation. In the future, we will recruit more students and conduct a longitudinal study to strengthen the robustness of the findings.